\documentclass[twocolumn,preprintnumbers,endnote,nofootinbib,prd]{revtex4}
\usepackage{graphicx}
\usepackage{amsmath}

\usepackage[hypertex]{hyperref}

\newcommand{\gsim}{\gtrsim}

\newcommand{\ord}[1]{\mathcal{O}{(#1)}}
\newcommand{\beq}{\begin{equation}}
\newcommand{\eeq}{\end{equation}}
\newcommand{\bea}{\begin{eqnarray}}
\newcommand{\eea}{\end{eqnarray}}

\newcommand{\ellpm}{\ell^+\ell^-}
\newcommand{\mP}{\bar{M}_{\rm P}}

\newcommand{\invfb}{${\rm fb}^{-1}$}
\def\l{\left}
\def\r{\right}
\def\misse{E\hspace{-0.2cm}/_{T}}

\begin{document}

\pagestyle{plain}

\title{\boldmath Warped Graviton ``$Z$ + Missing Energy" Signal at Hadron Colliders}

\author{Chien-Yi Chen
\footnote{email: cychen@bnl.gov}
}

\author{Hooman Davoudiasl
\footnote{email: hooman@bnl.gov}
}
\affiliation{Department of Physics, Brookhaven National
Laboratory, Upton, NY 11973, USA}

\author{Doojin Kim
\footnote{email: immworry@ufl.edu}
}
\affiliation{Institute for Fundamental Theory, Department of Physics,
University of Florida, Gainesville, FL 32611, USA}


\begin{abstract}

We examine the reach at hadron colliders for the lightest warped Kaluza-Klein (KK) graviton $G_1$ in the
$Z(\to \ellpm) Z(\to \nu{\bar \nu})$ channel, $\ell=e,\mu$, within Randall-Sundrum 
models of hierarchy and flavor where the Standard Model gauge fields and fermions propagate in the 5D bulk.  The reconstructed
$Z$ and the accompanying large missing energy allow for an efficient suppression
of backgrounds.  For reasonable parameters, a 
$\sim 2 \;(2.6)$~TeV $G_1$ can be discovered at 5$\sigma$ with $300\; \text{fb}^{-1} \,(3\;\text{ab}^{-1})$ of 14 TeV LHC data 
via our ``$Z$ + missing energy'' signal.  Using this signal, the discovery reach for $G_1$ at a future 100 TeV $pp$
collider is estimated to be as high as $\sim 10$~TeV.  We discuss mass determination of the singly produced $G_1$, 
using the energy distribution of its visible ($Z\to \ellpm$) decay product, by adapting a recently proposed method.
Based on our analysis, a mass measurement at the $\sim 5$\% level
with $\sim 3\;\text{ab}^{-1}$ of the 14 TeV LHC data can be feasible.

\end{abstract}
\maketitle


\section{Introduction}

The discovery of a Higgs boson at $\sim 126$~GeV, with properties very similar
to that of the Standard Model (SM), at the LHC \cite{Aad:2012tfa,Chatrchyan:2012ufa}
has been a significant step towards
uncovering the physics underlying electroweak symmetry breaking (EWSB).  At the same time,
conceptual questions regarding the Higgs mechanism, discussed at length
over the last few decades, can no longer be treated as hypothetical and take
on real urgency.  Of these, the stability of the $\sim 100$~GeV Higgs potential scale against
large quadratic corrections, known as the {\it hierarchy puzzle}, has been a main driver of
model building and its origin remains an open question.  Various ideas have been proposed over the
years to address this puzzle, ranging from non-trivial dynamics and low energy supersymmetry to models
with different kinds of extra dimensions, and many others.

An interesting resolution of the hierarchy is provided by the Randall-Sundrum (RS)
proposal \cite{Randall:1999ee}, where the background geometry is assumed to be a slice of AdS$_5$, {\it i.e.}
a 5D spacetime with constant negative curvature.
This 5D spacetime is characterized by curvature scale $k$ and is bounded by flat boundaries,
often called UV (Planck) and IR (TeV) branes.  Due to the
underlying {\it warped} geometry, UV brane scales of order $\mP \approx 2 \times 10^{18}$~GeV get redshifted
exponentially along the compact fifth dimension of size $\pi R$, with $R$ the
radius of compactification.  One would then get an IR brane scale of order $\mP \, e^{-k\pi R}\sim$~TeV for
$k R \approx 11$ which resolves the hierarchy without the need to introduce
very large numbers in the underlying 5D theory.  In the original RS model, all
SM fields were assumed confined to the IR brane.  Hence,
the main signals of this model were only from the gravitational
sector \cite{GW1,Davoudiasl:1999jd,GW2}, with a
distinct signal being the tower of Kaluza-Klein (KK)
gravitons \cite{Davoudiasl:1999jd}, appearing above the weak scale $\sim 100$~GeV.  The
masses of the lowest level gravitons are set by a scale of $\sim 1$~TeV,
similar to the scale of KK graviton coupling, the warped down $\mP$, at the IR brane.  Therefore,
the low lying KK states of the original model could in principle be produced as resonances, decaying into
all SM degrees of freedom at a basically universal rate.  In particular, one would have
striking signals in dileptons and diphotons, reconstructing to TeV-scale spin-2
resonances in this scenario.

Soon after the introduction of the RS proposal, it was realized that the hierarchy solution, while
requiring the Higgs field to be localized near the IR brane, does not exclude the possibility that
other fields may propagate in the
bulk \cite{Goldberger:1999wh}.  In fact, placing
the SM gauge fields \cite{Davoudiasl:1999tf,Pomarol:1999ad} and fermions \cite{Grossman:1999ra}
in all five dimensions was shown to lead to
a realistic picture of flavor \cite{Gherghetta:2000qt}.
This was achieved through introduction of 5D masses for fermions \cite{Grossman:1999ra},
allowing for light (heavy) fermions to have zero mode profiles localized away (towards)
the Higgs on the IR brane.  In this way, the same warped background could explain
both the hierarchy and flavor puzzles of the SM while being enriched with new signatures
associated with the additional KK states of the gauge and the fermion fields.  However, in these
new warped models of hierarchy and flavor, the KK modes mainly couple to heavy SM degrees of
freedom: $t$, $W^\pm$, $Z$, and the Higgs.  In particular, the KK graviton
couplings to light fermions become negligible while their couplings to massless photons and
gluons are suppressed by a volume factor $\sim 1/(k\pi R)$ \cite{Davoudiasl:2000wi}.  Consequently, unlike in
the original RS model, the main decay modes of KK gravitons are now the heavy fields
that do not yield simple final state signals.

The key aspects of collider phenomenology in the context of warped models
of hierarchy and flavor (bulk SM) have been studied in a number of works; for a sample, see
Refs.~\cite{Davoudiasl:2000wi, Agashe:2006hk,Lillie:2007yh,Fitzpatrick:2007qr,Agashe:2007zd,Lillie:2007ve,Djouadi:2007eg,
Agashe:2007ki,Antipin:2007pi,Agashe:2008jb,
Davoudiasl:2009cd}.  In particular, the prospects for
detecting the lightest KK graviton $G_1$ were studied in Ref.~\cite{Fitzpatrick:2007qr} in the $G_1\to t{\bar t}$
channel, in Ref.~\cite{Agashe:2007zd} for $G_1\to Z_LZ_L$, i.e., the decay into longitudinal $Z$ pairs with both $Z$ bosons decaying leptonically, $Z\to \ell^+\ell^-$ for $\ell=e, \mu$, while Ref.~\cite{Antipin:2007pi} examined the $G_1\to W_LW_L$ signal.  These  studies suggest that the reach for $G_1$ is at most $\sim$ 2~TeV with
$\sim {\rm few}\times 100$~\invfb of data at the 14~TeV LHC.\footnote{The relevant matrix element in Ref.~\cite{Agashe:2007zd} should include another factor of 1/2 and the corresponding cross section for $gg\rightarrow VV$ should be reduced by 1/4.  Therefore, the reach for KK gravitons in Ref.~\cite{Agashe:2007zd} is lowered from the $\sim$ 2 TeV projection in that work.}

In this work, we extend the study in Ref.~\cite{Agashe:2007zd}
to the case of $G_1 \to Z_L(\to \ell^+\ell^-) Z_L(\to \nu {\bar \nu})$,
which can be taken as complementary work.
Given that branching fraction of the $Z$ into neutrinos is roughly a factor of 3 larger than that into $\ell^+\ell^-$,
one could expect this channel to provide an important search mode.
The price of having a larger branching ratio is that the relevant final state involves invisible neutrinos and is hence 
not fully reconstructible. Nevertheless, the invisible neutrinos provide an important handle to reduce the relevant background. The reason is that the KK graviton $G_1$ is typically expected to be heavy so that its subsequent $Z$ boson decay products are likely to have large transverse momenta and boosts. Therefore, as we will show, the large missing energy and also the small distance between the two final state leptons in this case can be used to cut the background efficiently.  We will consider integrated luminosities of 300 $\text{fb}^{-1}$ and 3 $\text{ab}^{-1}$ at the 14 TeV LHC.   
Also, we will estimate the reach at a future 100 TeV proton collider
with 1 and 3 $\text{ab}^{-1}$ of data.\footnote{For other studies of RS models at future colliders, 
see also Refs.~\cite{Agashe:2013fda,Agashe:2013kxa}.} While the reach at the LHC is 
about $2-3$ TeV, the 100 TeV machine has the potential to discover a $\sim$ 10 TeV KK graviton.

As mentioned before, our signal cannot be fully reconstructed to yield the mass $m_{G_1}$ of $G_1$, due to the loss of kinematic
information about the neutrinos that escape the detector.  However,  KK gravitons of interest in our work are generally heavy, 
$m_{G_1}\gsim 2$~TeV, and thus mostly produced at rest at the LHC.  Also, initial state gluon parton distributions 
are symmetric and are less likely to lead to a boosted KK graviton.  This suggests that each $Z$ in the decay will 
carry approximately $m_{G_1}/2$ of energy, and therefore
the distribution of the visible energy (from $Z\to \ellpm$) can be used to estimate the value of $m_{G_1}$.  
To this end, we adapt the methodology proposed in Ref.~\cite{Agashe:2012bn} 
for determining the mass of a resonance that undergoes a 2-body decay, using  
partial kinematic information.  The results in Ref.~\cite{Agashe:2012bn} imply that under certain conditions, 
the peak in the energy distribution of a visible `massless' daughter particle equals 
its energy in the rest frame of the mother state.\footnote{The method is relevant even to the case where the two decay products can be fully reconstructed because we may ignore one of them.  In such a case, resonance reconstruction is typically a better choice, but the proposal in Ref.~\cite{Agashe:2012bn} can become competitive if one of the decay products is difficult to reconstruct.}  
Although some of the requisite conditions
are not strictly met for our process, we  find that the energy distribution ansatz implied by the 
discussion in Ref.~\cite{Agashe:2012bn} can yield a $\sim 5\%$ measurement of $m_{G_1}$, given a statistical sample of 
$\mathcal{O}(100)$ events.  
    
In the next section, we will outline the setup and the range of parameters considered in this work. We then move onto the discussion about the search strategy for our signal in Section III. Section IV contains our results on the reach of the 14 TeV LHC and a 100 TeV collider for the lightest KK graviton. In Section V, we discuss the prospects for determining the KK graviton mass using the 
final state of interest in our work.  The last section is reserved for a brief discussion and our conclusions.

\section{Setup and Parameters}

The models we consider could be endowed with further symmetries to control
unwanted deviations from precision EW data \cite{custodial}, allowing the lightest
gauge KK states to have a mass of order 2-3~TeV \cite{Carena:2007ua}.  However,
we will mostly follow the assumptions used in Ref.~\cite{Agashe:2007zd}
and focus on a simple realization of warped hierarchy and flavor models.  In particular,
we assume that the bulk gauge field content is that of the SM and the only fermion
with significant coupling to $G_1$ is the right-handed top quark $t_R$.  These assumptions capture the main
features of phenomenology and the results we find can be useful estimates for somewhat more
complicated models.  An important parameter governing warped KK graviton phenomenology is the ratio
$c\equiv k/\mP$.  The size of this parameter is a measure of whether the underlying
RS background can be well-described by the Einstein-Hilbert action in general relativity and strong gravity
effects can be ignored; typically this is the case for $c$ not much larger than order unity.  Here, we will consider
values $c\leq 2$, which could be consistent with a classical
description of the geometry \cite{Agashe:2007zd}.

We note that in the RS background, the masses of the gauge field KK states start at $2.45 \times k e^{-k \pi R}$ while
the KK graviton tower starts at $3.83 \times k e^{-k \pi R}$. Hence, we expect the lightest
KK graviton to be roughly a factor of 1.6 heavier than the lightest gauge KK state such as the lightest KK gluon. The lightest KK gluon, with its
coupling to quarks proportional to the strong coupling constant, typically provides the best limit on warped models with bulk flavor and is currently constrained to be heavier than $\sim 2$~TeV by the LHC data \cite{ATLAS-KKgluon,CMS-KKgluon}. Hence, the simplest RS-type models would then predict
that the lightest KK graviton should lie above $\sim 2.5-3$~TeV.

Here, we would like to add that warped KK phenomenology can be significantly
different in volume ($k R$) truncated ``Little RS'' models, leading to enhanced collider
signals \cite{LRS,LRSLHC}.  In those models, a smaller hierarchy between the TeV scale and
a large but sub-Planckian scale (for example, associated with flavor physics) is addressed.
In what follows, we only consider warped models that explain the Planck-weak hierarchy.

\section{SM background and selection criteria}

Our signal events are defined by the singly-produced KK graviton decaying into a pair of $Z$ gauge bosons, one of which decays further leptonically while the other decays invisibly:
\bea
pp \rightarrow G_1 \rightarrow ZZ \rightarrow \ell^+\ell^-\nu \bar{\nu}
\eea
where $\ell$ includes $e$ and $\mu$ only. Due to this signal process, the relevant collider signature is characterized by the opposite-signed same flavor di-lepton and large missing energy. Therefore, the relevant SM backgrounds can be $Z+\nu \bar{\nu}$ in which $Z$ decays leptonically, $W^{\pm}Z$ in which $Z$ decays leptonically while $W^{\pm}$ decays into a lepton and its corresponding neutrino,
and $W^+W^-$ in which the two $W$ gauge bosons decay leptonically.
For the last case, even though it has a large production cross section, we can suppress the 
background by requiring opposite-signed same flavor leptons whose invariant mass falls into the $Z$ mass window.

In order for $W^{\pm}Z$ to appear as a background, the $W$ gauge boson should decay leptonically, and the relevant lepton ($e$ or $\mu$) should be missed, which happens only a small fraction of the time for the
typical lepton transverse momentum $p_T^{\ell}$ in this case.  However, if the lepton is missed due to small $p_T^{\ell}$, with  the missing transverse energy $\misse$ dominantly from $\nu$, the background can still be suppressed by requiring a hard $\misse$ cut.  Another way of getting a background event from this channel is for $W^{\pm}$ to decay into the $\tau^{\pm}$ which decays hadronically to give rise to soft jet(s) along with another neutrino resulting in additional missing momentum. However, our simulation studies suggest that the contribution from this process should be subdominant with the aid of the set of cuts to be described shortly.

\begin{figure}[t]
\centering
\includegraphics[width=8.4cm,height=6.3cm]{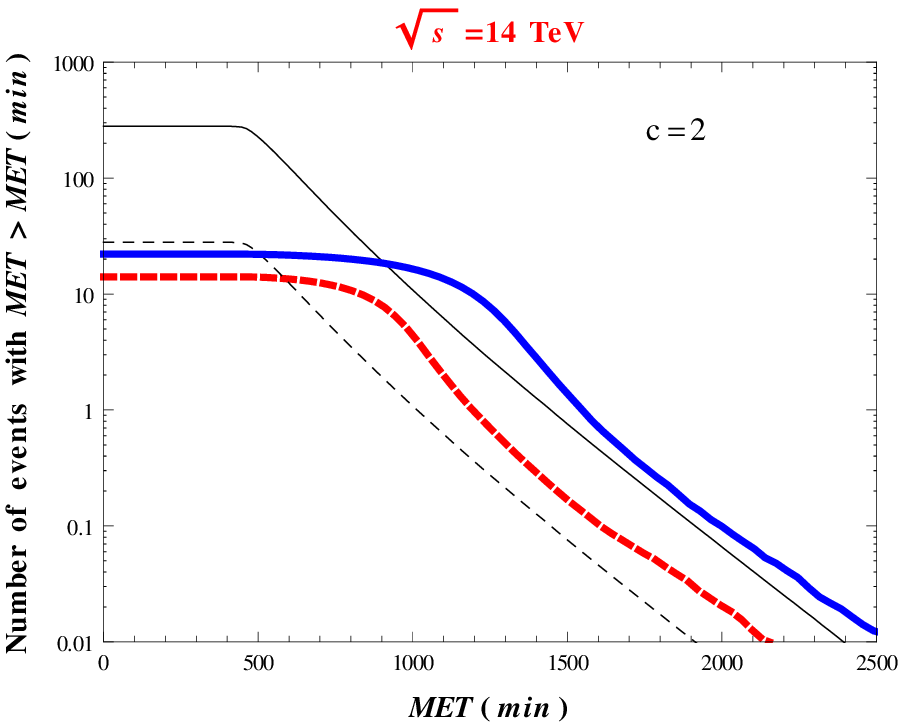}
\caption{Distributions of the number of signal and the dominant SM background events that pass the selection criterion $\misse > \misse$(min) for set ({\it i}) (red thick dashed line) and set ({\it iii}) (blue thick solid line) in TABLE~\ref{tab:reach300} at $\sqrt{s}=14$ TeV and $c = 2$. The thin lines represent backgrounds for integrated luminosities of 300 $\text{fb}^{-1}$ (black dashed) and 3 $\text{ab}^{-1}$ (black solid). The basic cuts in Eq.~(\ref{eq:basiccuts}) and $\Delta R_{\ell\ell}<0.4$ are applied. \label{fig:MET14} }
\vspace{0.5cm}
\includegraphics[width=8.4cm,height=6.3cm]{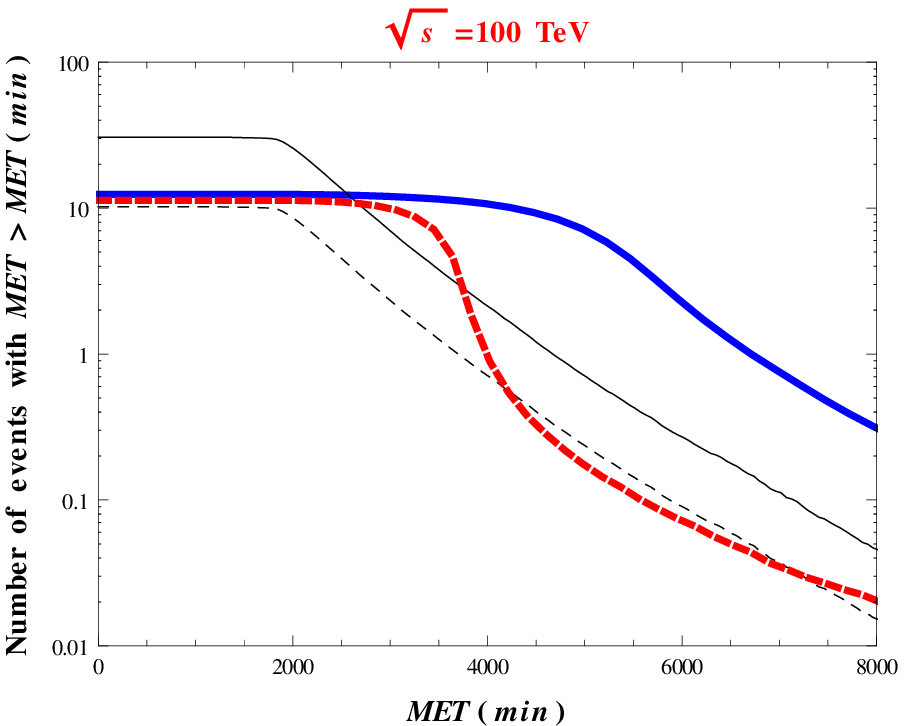}
\caption{Distributions of the number of signal and the dominant SM background events that pass the selection criterion $\misse > \misse$(min) for set ({\it v}) (red thick dashed line) and set ({\it vi}) (blue thick solid line) in TABLE~\ref{tab:reach300} at $\sqrt{s}=100$ TeV for $c = 1$ and $c = 2$, respectively. The thin lines represent backgrounds for integrated luminosities of 1 $\text{ab}^{-1}$ (black dashed) and 3 $\text{ab}^{-1}$ (black solid). The basic cuts in Eq.~(\ref{eq:basiccuts}) and $\Delta R_{\ell\ell}<0.1$ are applied. \label{fig:MET100} }
\end{figure}

For simulating the signal and background events at the parton level, we employ the matrix element generators $ \texttt{CalcHEP3}$~\cite{Belyaev:2012qa} and $\texttt{MadGraph5}$~\cite{Alwall:2011uj}, respectively, taking the parton distribution functions of $\mathtt{CTEQ6L1}$~\cite{Pumplin:2002vw}.
For the event {\it pre}-selection, we basically follow a similar strategy to that taken in Ref.~\cite{Aaltonen:2011jn} along with slight modifications:
\bea
&p_T^{\ell}>25 \text{ GeV},\; |\eta^{\ell}|<2.4,\;\misse >400 \, (1500) \,\hbox{ GeV}& \nonumber \\
&66 < m_{\ell\ell} < 116 \text{ GeV}\,,& \label{eq:basiccuts}
\eea
where $\eta^{\ell}$ denotes the pseudo-rapidity of $\ell=e, \mu$ and the larger $\misse$ cut in the parentheses is for $\sqrt{s}= 100$~TeV.
Beyond these pre-selection cuts, we find that additional 
cuts on $\misse$ and $\Delta R_{\ell\ell}\equiv \sqrt{(\Delta \phi^{\ell})^2  + (\Delta \eta^{\ell})^2}$, with $\phi^\ell$ denoting the azimuthal angle, play a key role in rejecting the background events further. Since the KK graviton is expected to be quite heavy, our signal typically gives rise to large missing momentum.  
To optimize the $\misse$ cut one can plot the distributions of the number of signal and the dominant SM background events
that pass the selection criterion $\misse > \misse(\rm{min})$ as a function of $\misse(\rm{min})$, as shown in FIGs. 1 and 2.
Note that the events are selected after applying the basic cuts listed in Eq.~(\ref{eq:basiccuts}) along with a minimal choice of $\Delta R_{\ell\ell}$ cut which will be discussed shortly.
The advantage of these distributions is that they help in finding a $\misse$ 
cut that would optimize statistical significance, i.e., $S/\sqrt{B}$.
We will explain this in detail in the next section.
These distributions suggest that a harder $\misse$ cut can suppress a large fraction of background events, and therefore significantly improve 
the signal-to-background ratio.

\begin{figure}[t]
\centering
\includegraphics[scale=0.65]{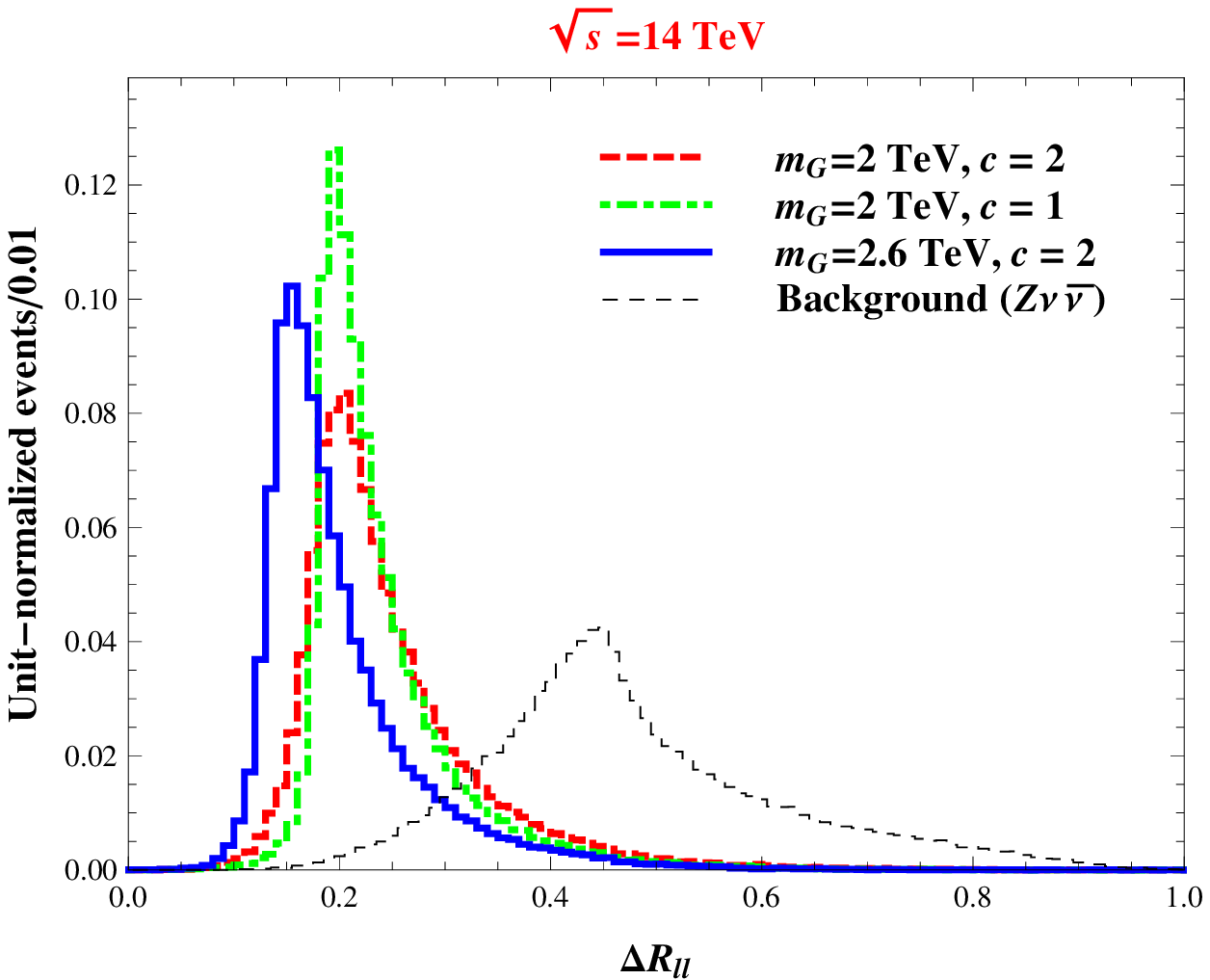}
\caption{\label{fig:DRCut14TeV} The $\Delta R_{\ell\ell}$ sensitivity of signal [sets ({\it i}), ({\it ii}), and ({\it iii}) in TABLE~\ref{tab:reach300}] and background events with the collider of $\sqrt{s}=14$ TeV after imposing the cuts in Eq.~(\ref{eq:basiccuts}).} \vspace{0.5cm}
\includegraphics[scale=0.65]{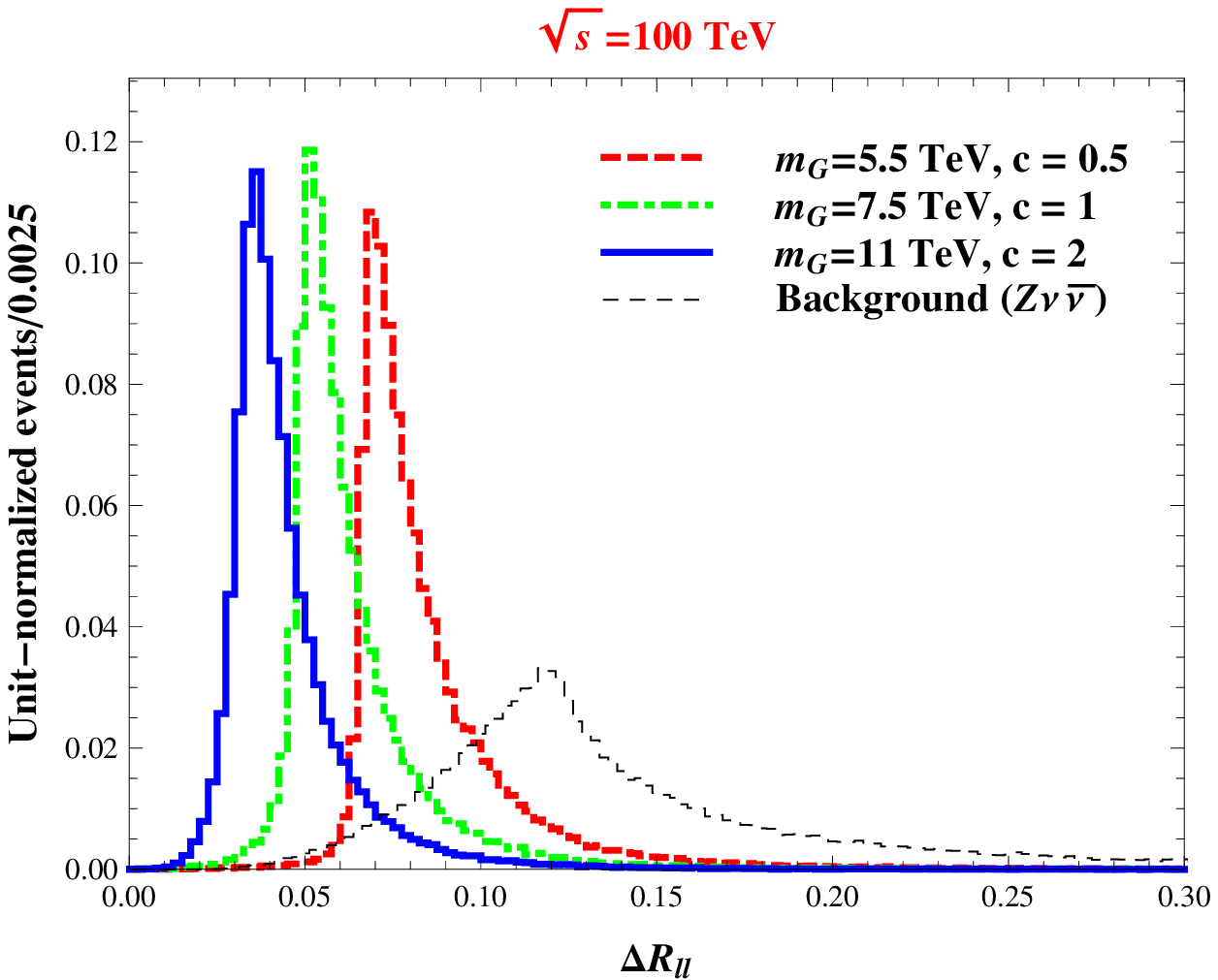}
\caption{\label{fig:DRCut100TeV} The $\Delta R_{\ell\ell}$ sensitivity of signal [sets ({\it iv}), ({\it v}) in TABLE~\ref{tab:reach300}), and ({\it vi})] and background events with the collider of $\sqrt{s}=100$ TeV after imposing the cuts in Eq.~(\ref{eq:basiccuts}).}
\end{figure}

This observation also motivates a small $\Delta R_{\ell\ell}$ cut because the highly boosted $Z$ leads to two collimated decay products, $\ell^+$ and $\ell^-$. On the other hand, most of the $Z$ gauge bosons in the background at hand are produced nearly at threshold and hence are not highly boosted. Therefore, the separation between the two leptons from the decay of such $Z$ bosons is expected to be (relatively) large for most of the background events. In this sense, a small $\Delta R_{\ell\ell}$ can reduce the number of such background events considerably. FIGs.~\ref{fig:DRCut14TeV} and~\ref{fig:DRCut100TeV} compare unit-normalized distributions of the benchmark signal events in TABLE~\ref{tab:reach300} and the $Z+\nu\bar{\nu}$ background events in $\Delta R_{\ell\ell}$ with $\sqrt{s}$ being 14 TeV and 100 TeV, respectively. The events are selected after the pre-selection cuts enumerated in Eq.~(\ref{eq:basiccuts}). Note that for the 100 TeV case, we adjust the minimal $\misse$ cut in Eq.~(\ref{eq:basiccuts}) to be $\misse>1500$ GeV. Certainly, we observe that a small $\Delta R_{\ell\ell}$ is favored for the signal events, from which we are led to the following $\Delta R_{\ell\ell}$ cuts:
\bea
\begin{array}{l l}
\Delta R_{\ell\ell} <0.4 & \hbox{for }\sqrt{s}=14\hbox{ TeV} \\
\Delta R_{\ell\ell} <0.1 & \hbox{for }\sqrt{s}=100\hbox{ TeV}. \label{eq:DRcut}
\end{array}
\eea
Note that $\Delta R_{\ell\ell}$ decreases as the mass of $G_1$ increases.  Thus, for low 
values of $m_{G_1}$, it would be more difficult to discriminate the signal from the SM background.

\section{Discovery potential}

In this section, we discuss the discovery potential of the KK graviton based upon the cuts in Eqs.~(\ref{eq:basiccuts}) and~(\ref{eq:DRcut}) together with a suitable choice of the $\misse$ cut depending on the study point of interest. TABLE~\ref{tab:reach300} lists six representative cases that are used in our simulations and the corresponding statistical significance along with the optimized $\misse$ and $\Delta R_{\ell\ell}$ cuts for various masses and $c$ parameters of the KK graviton.

We begin with the case of $\sqrt{s}=14$ TeV with integrated luminosities of 300 $\text{fb}^{-1}$ for set ({\it i}) and 3 $\text{ab}^{-1}$ for sets ({\it ii}) and ({\it iii}). The selection of the appropriate $\misse$ cut can be understood by observing FIG.~\ref{fig:MET14}. For example, we can find that the best $\misse$ cut for set ({\it i}) that leads to the largest $S/\sqrt{B}$ arises around 0.9 TeV while for set ({\it iii}), it does around 1.2 TeV. Furthermore, we can also see that these cuts are all quite close to half the mass of $G_1$. This is not surprising, as the typical KK graviton under consideration is so heavy that it is likely to be produced nearly at rest. Since each KK graviton decays into {\it two} $Z$'s which are effectively massless, the typical energy scale of each decay product is $\sim m_{G_1}/2$ in such a situation. Given that the $Z$ decaying into neutrinos is the main source of $\misse$, the above-mentioned best $\misse$ cuts can be expected.  
From this argument one could claim that both sets ({\it i}) and ({\it ii}) should have the same $\misse$ cut because the mass of $G_1$ is the same for these two benchmark points. However, in TABLE~\ref{tab:reach300} set ({\it i}) has a slightly softer $\misse$ cut than that of set ({\it ii}). The reason is that the relevant discovery here is defined such that the number of the signal events has to be at least 10 
concurrently with usual $S/\sqrt{B}\geq5$.  We have assumed 100\% efficiency, as a fair approximation for the clean leptonic signals.

One noteworthy observation is that the ratio of the number of the signal events of set ({\it i}) (denoted as $S_i$) to that of set ({\it ii}) (denoted as $S_{ii}$) is about 0.4. This number can be understood
from Eqs.~(9) and~(10) of Ref.~\cite{{Agashe:2007zd}}: the parton-level cross section is proportional to the square of the amplitude
\beq
\left|{\cal M}^{G_1}\left( g g \rightarrow VV \right)\right|^2 \propto
\left|\frac{(x_{G_1} c)^2}
{ \hat{s} - m_{G_1}^2 + i\Gamma_G m_{G1} }\right|^2
\label{MG1}
\eeq
where $V=W,Z$ and 
\bea
\Gamma_{G_1} & = & \frac{ 13 (c\, x_{G_1})^2 \, m_{G_1} }{ 960 \pi}.
\eea
For small $\Gamma_{G_1}/m_{G_1}$, using the narrow width approximation~\cite{Han:2005mu} one can find that
\bea
 \frac{1}{(\hat{s}-m_{G_1}^2)^2+\Gamma_{G_1}^2 m_{G_1}^2}
\approx \frac{\pi}{\Gamma_{G_1} m_{G_1}}\delta(\hat{s}-m_{G_1}^2).
\eea
As a result, when the width of $G_1$ becomes very narrow compared with $m_{G_1}$, the total parton-level cross section is proportional to $c^2$.  
Thus, the ratio of $S_i$ to $S_{ii}$ is as follows:
\bea
\frac{S_i}{S_{ii}}=\frac{{\cal L}_i}{{\cal L}_{ii}} \left(\frac{c_i}{c_{ii}}\right)^2= 0.4.
\eea
In TABLE~\ref{tab:reach300}, this ratio is not exactly equal to 0.4 because $\Gamma_{G_1} / m_{G_1}$ may not be still small enough or the cuts on $\misse$ are slightly different. Nevertheless, this argument provides a good intuition on the expected number of signal events for $G_1$ of interest with different $c$ parameters.

\begin{table}[h]
\centering
\caption{\label{tab:reach300} Statistical significance (i.e., $S/\sqrt B$) of six sets of benchmark points for $\sqrt s =14$ and 100
TeV. Also, the number of signal events is required to be at least 10. $\misse$ and $\Delta R_{\ell\ell}$ cuts are optimized for each benchmark point to obtain the best significance. }
\begin{tabular}{c|c c c|c c c}
\hline
 & ({\it i})& ({\it ii}) & ({\it iii}) & ({\it iv}) & ({\it v})& ({\it vi}) \\
\hline \hline
$\sqrt s $ (TeV) & 14&14&14&100&100&100 \\
\hline
${\cal L}$ ($\text{ab}^{-1}$)  & 0.3& 3& 3& 1& 1& 3 \\
\hline
$c=k/\bar{M}_P$ & 2&1&2& 0.5&1&2\\
\hline
$m_{G_1}$ (TeV)  & 2& 2& 2.6& 5.5& 7.5& 11 \\
\hline
$\misse$ cut (TeV)  & $>$0.8 &$>$0.9 &$>$1.2 &$>$2.4 &$>$2.9 &$>$3.7  \\
\hline
$\Delta R_{\ell\ell}$ cut  & $<$0.4 &$<$0.4 &$<$0.4 &$<$0.1 &$<$0.1 &$<$0.1  \\
\hline
S  & 10& 28& 11& 12& 10& 11 \\
\hline \hline
$S/\sqrt B$  & 5.7& 6.4& 5.4& 5.1& 6.2& 6.5 \\
\hline

\end{tabular}
\end{table}

Next, we move our attention to the numerical results for a future proton-proton collider with $\sqrt{s}=100$ TeV and integrated luminosities of 1 $\text{ab}^{-1}$ for sets ({\it iv}) and ({\it v}) and 3 $\text{ab}^{-1}$ for set ({\it vi}). The relevant analysis can be performed in an analogous way to the case of $\sqrt{s}=14$ TeV. Our simulation study suggests that the discovery potential can reach $m_{G_1} = 5.5$ TeV or higher depending on the values of the $c$ parameter and the luminosity that the machine can acquire. Since $m_{G_1}$  is very large, the associated $\misse$ cut also becomes even harder while the separation between the two visible leptons becomes closer as seen from FIG.~\ref{fig:MET100}.

The discovery potential looks promising for the 14 TeV LHC; even with a luminosity of 300 $\text{fb}^{-1}$, the discovery reach is around $m_{G_1} = 2$ TeV with a statistical significance of 5.7 $\sigma$, assuming $c\approx 2$ .  Since the statistical significance increases as $\sqrt{\mathcal{L}}$, larger amounts of data allow reaching higher masses and smaller values of $c$ (the cross section is proportional to $c^2$).  We observe that for an integrated luminosity of 3 $\text{ab}^{-1}$, the discovery potential can reach  $m_{G_1} = 2$ and $m_{G_1} = 2.6$ TeV for $c=1$ and $c=2$, respectively.

For KK gravitons with higher masses or smaller $c$ parameters, such as $c=0.5$, enhanced cross sections at a more energetic machine would be required. For example, at  a 100 TeV hadron collider, we find that discovering a KK graviton of $m_{G_1}=11$ TeV with $c=2$ is possible, for an integrated luminosity of 3 $\text{ab}^{-1}$ [see set ({\it vi}) in TABLE~\ref{tab:reach300}].  As another example, set ({\it iv}) represents a special case where $c = 0.5$, which is the smallest among the benchmark points in TABLE~\ref{tab:reach300}.  As mentioned above, the cross section is proportional to $c^2$ so that the production rate for this point is  suppressed and this makes the discovery challenging. However, our results show that with a luminosity of 1 $\text{ab}^{-1}$ the reach of the 100 TeV machine for this point is $m_{G_1} =5.5$ TeV with a statistical significance of 5.1$\sigma$.

\section{Prospects for Mass Measurement \label{sec:massmeasurement}}

Once the KK graviton $G_1$ is discovered, one could attempt to measure its mass.  As discussed before, the approximate value of the KK graviton mass can be inferred from the (reconstructible) $Z$ energy distribution, since $G_1$ is typically produced nearly at rest at the LHC and hence each decay product has $\sim m_{G_1}/2$ of energy.  However, for the cases with low statistics and non-negligible background one may need to apply well-suited methods to optimize the mass determination. For this purpose, we adapt the novel proposal in Ref.~\cite{Agashe:2012bn}, based on the observation that, in a 2-body decay, 
the laboratory frame energy distribution of a (massless) visible daughter 
particle peaks at the value of its energy in the rest frame of the mother particle.  Based on general arguments,  Ref.~\cite{Agashe:2012bn} 
proposed an ansatz describing such an energy distribution, which can be used to determine the peak more accurately.

We apply a method akin to that discussed above for identifying the location of the peak in the $Z$ energy distribution, and hence estimating $m_{G_1}$.  
However, we note that the ansatz proposed in Ref.~\cite{Agashe:2012bn} was focused on resonances produced in pairs, generally yielding  
particles with non-zero boosts.  In our case though, the KK graviton is singly produced and typically with negligible boost. 
As described in Ref.~\cite{Agashe:2012bn}, the peak in the former case corresponds to an extremum, while 
in the latter case the peak corresponds to a cusp.  Suppose that the $Z$ energy, $E_Z$, in the laboratory frame is distributed according to 
$f(E_Z)$, and the rest-frame energy of $Z$ is denoted by $E_Z^*$. To capture the cusp feature of the peak in $f(E_Z)$ for the case of interest here, 
we adopt a new ansatz that is expected to accommodate the cusp structure around the peak (i.e., $E_Z=E_Z^*$) \cite{Franceschini}:
\bea
f(E_Z)=N \exp\l[-w\left(\frac{E_Z}{E_Z^*}+\frac{E_Z^*}{E_Z}-2\right)^q \r]\,.\label{eq:ansatz1}
\eea
The first derivative of $f(E_Z)$ behaves as 
\bea
f'(E_Z)
\propto \text{sgn}(E_Z-E_Z^*) \exp\l[-w\l(\frac{E_Z}{E_Z^*}+\frac{E_Z^*}{E_Z}-2\r)^q \r], 
\eea
where $q\in (0,1)$ encodes the steepness around the peak position, $E_Z=E_Z^*$.  
The sign function guarantees discontinuity in the slope at $E_Z=E_Z^*$, which yields a cusp in $f(E_Z)$ at $E_Z^*$.

The other possible issue is that the method proposed in Ref.~\cite{Agashe:2012bn} assumes a flat angular distribution 
from a scalar or else an unpolarized production process.  However, for our KK graviton, a tensor particle, 
the decay product $Z$ has an angular preference relative to the direction of the incoming partons ($\sim \sin^4\theta$, 
in the parton center of mass frame)~\cite{Agashe:2007zd}. Nevertheless, our simulations suggest that the departure, in our case, 
from the assumptions leading to Eq.(\ref{eq:ansatz1}) do not 
affect the utility of the formalism significantly.  As KK gravitons are produced nearly at rest, the energy of 
$Z$ final states is centered around $m_{G_1}/2$ and we may apply the above ansatz to locate the peak.  

\begin{figure}[t]
\centering
\includegraphics[scale=0.65]{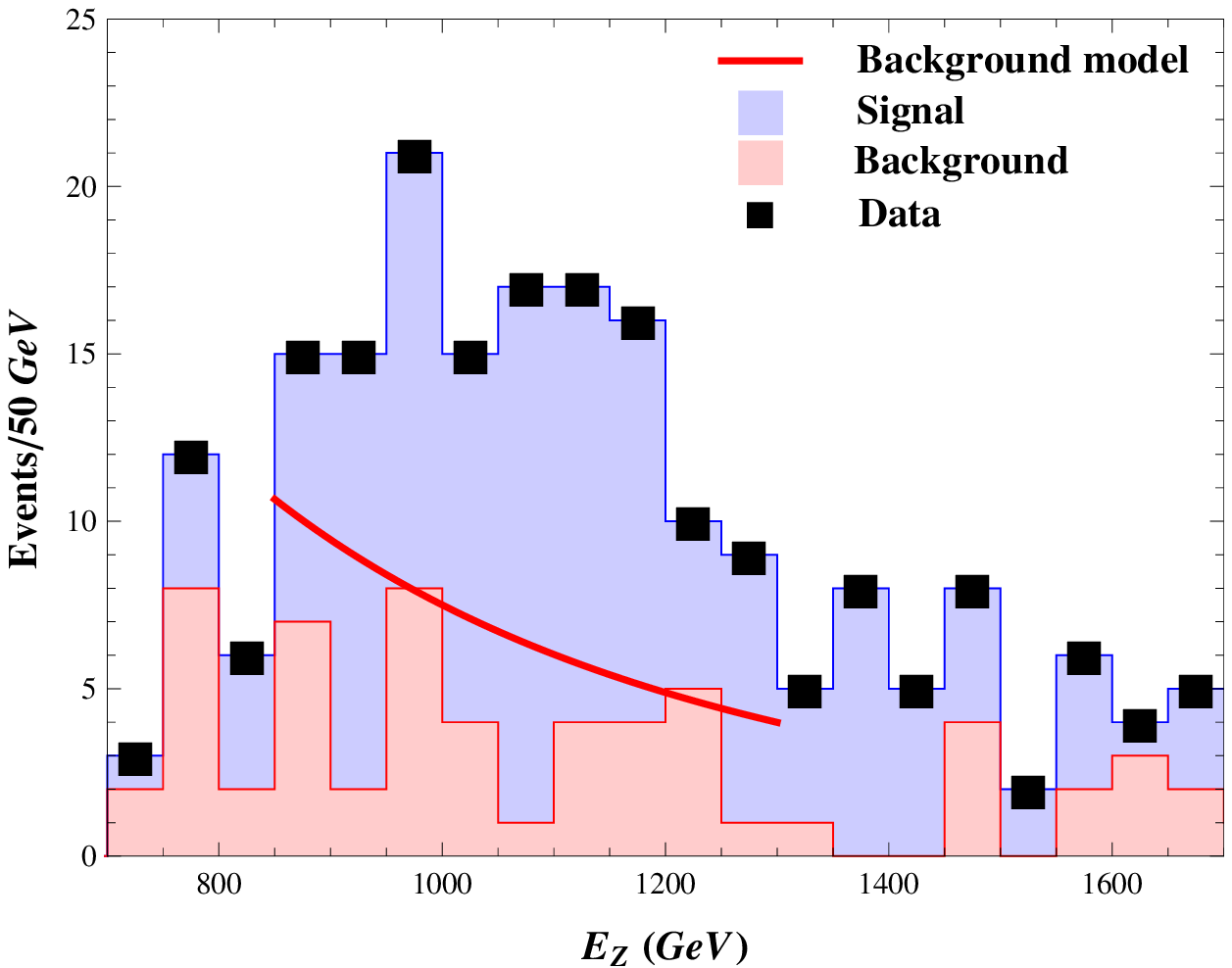}
\includegraphics[scale=0.65]{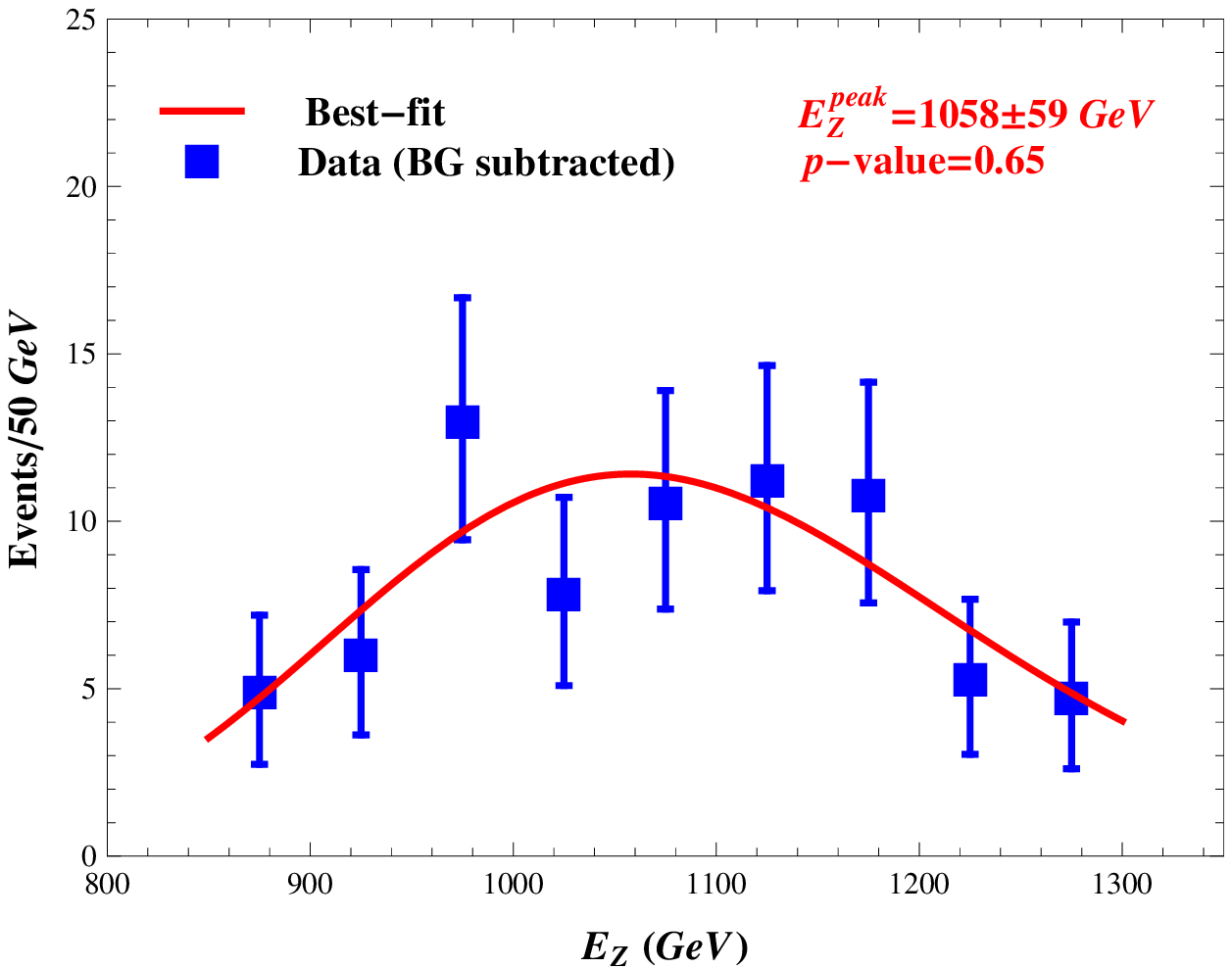}
\caption{\label{fig:samplefit} A sample $Z$ energy distribution
including background events (top panel) and the fit of the data to Eq.~(\ref{eq:ansatz1}),
after background subtraction (bottom panel). }
\end{figure}

For the purpose of demonstrating the relevant technical details, we take $m_{G_1}=2$~TeV and $c=2$, for $G_1$ 
produced at the 14 TeV LHC, which according to TABLE~\ref{tab:reach300} can be discovered with $300\;\text{fb}^{-1}$ of data.
We assume 3 $\text{ab}^{-1}$ of integrated luminosity to secure the necessary statistics. Also, we relax the $\misse$ cut a little ($\misse >0.7$ TeV) in order to avoid significant deformation of the shape in the vicinity of the expected peak position. With this luminosity and setup ~$\sim150$ events are expected to be detected, with a signal strength (i.e., $S/B$) of $\sim 2.5$. Following Ref.~\cite{Agashe:2013eba}, we subtract the background in the fitting region with a proper background template 
\bea
f_{\text{BG}}(E)=N_{\text{BG}}\exp\l(-w_{\text{BG}}\sqrt{E} \r) \label{eq:BGtemp}\,,
\eea
where $N_{\text{BG}}$ and $w_{\text{BG}}$ are the normalization and the model parameters, respectively. 
Our simulation suggests (top panel in FIG.~\ref{fig:samplefit}) 
that the background can be fairly well-described by this template in the associated fitting range.

For the signal fit, we consider an ensemble of 200 pseudo-experiments with the data corresponding to the 3 $\text{ab}^{-1}$ of the integrated luminosity at $\sqrt{s}=14$ TeV as mentioned before. For each pseudo-experiment, we perform a Log-Likelihood fit with the template given in Eq.~(\ref{eq:ansatz1}). We only take the data approximately within the half maximum, that is, [850,\;1300] GeV where most of the signal events are populated. We also found that $q\sim0.9$ describes the kink structure around the peak well enough. The extracted peak position and its error estimation
at 95\% confidence level (C.L.) over 200 pseudo-experiments with a bin size of 50 GeV are given by
\bea
\langle E_Z^{\text{peak}} \rangle =1052 \pm 53 \text{ GeV}
\eea
with a mean $p$-value of 0.17 for our fit. FIG.~\ref{fig:samplefit} demonstrates a sample fit which would be performed in the actual experiment. The top panel shows the $Z$  energy distribution given by the combination between the signal (blue area) and the background (red area) events. The red curve denotes the relevant background model given in Eq.~(\ref{eq:BGtemp}), for which the model parameters have been extracted by the fit of the background event sample of large statistics. 
We see that it is fairly close to the actual background even with the small number of events. After subtracting the full distribution by such a background model in the above-mentioned fitting range, we fit the output, and the fit result is shown in the bottom panel, where the red solid curve represents the best-fit.

The distributions in the best-fit $E_Z^{\text{peak}}$ and its error at 95\% C.L. are symmetric around their central values, and no special features arise. The peak position, which is the same as the rest-frame energy of $Z$, can be translated into the mass of the KK graviton: 
\bea
E_Z^{\text{peak}}=\frac{m_{G_1}}{2}
\eea
from which we infer the mass parameter of interest 
\bea
\langle m_{G_1} \rangle = 2104\pm 106 \text{ GeV}.
\eea
The measured value is fairly precise and accurate: the corresponding error is small, i.e., $\sim 5\%$, and the central value is in a good agreement with the input value at the 95\% C.L. In addition, the associated $p$-value is acceptable. Given this result, we expect that KK graviton mass measurement, using 
our final state, will be feasible for interesting parameter ranges at the 14 TeV LHC, given $\sim 3$ $\text{ab}^{-1}$ of the integrated luminosity  

Before closing this section, we remark on a couple of potential issues. The first one is that as pointed out in Ref.~\cite{Agashe:2007zd}, the decay width of the KK graviton is no longer negligible once the associated $c$ parameter becomes large. As a consequence, the $Z$ energy distribution is {\it not} $\delta$-function-like even in the rest frame of the KK graviton, i.e., it develops a non-trivial distribution peaked at the $E_Z^*$ evaluated with the nominal mass of the KK graviton.   Remarkably, our simulation suggests that such effects would not make a significant impact on the final output, i.e., the approach introduced above can be a good approximation with respect to measurement of $m_{G_1}$.

Another point regarding the above mass measurement approach is comparison with those using 
$p_T$ distribution of $Z$ gauge bosons. Since the KK gravitons are singly produced, the $p_T$ distribution of the visible 
$Z$ gives rise to the famous Jacobian peak, which 
could also give a handle on the value of $m_{G_1}$. However, given a non-negligible decay width like the example studied here, the sharpness of such a peak gets reduced so that the identification of the peak in the $p_T$ distribution may not be straightforward. Moreover, locating the peak position in the $p_T$ distribution becomes more challenging in the presence of initial state radiation (ISR) which can shift the peak in $p_T$.  If the non-trivial angular 
distribution of the final state $Z$ ($\propto \sin^4\theta$) can be roughly ignored, the arguments in Ref.~\cite{Agashe:2012bn} imply that as KK gravitons 
get boosted due to effects of ISR, the location of the energy peak should not be affected.\footnote{Final state radiation (FSR) can shift the peak in the energy distribution because it effectively results in three-body decays for which the 
method in Ref.~\cite{Agashe:2012bn} is not applicable~\cite {oai:arXiv.org:1212.5230}.  
However, for the signal channel under consideration, the effect from FSR is negligible due to the leptonic final states.}  
Then, the energy variable in our analysis could yield a good estimate of $m_{G_1}$ 
even when more realistic details are taken into account.  However, a more detailed simulation is required to examine this question, which is outside the 
scope of our work.  

\section{Conclusions}

In this work, we studied the discovery potential for the lightest warped (Randall-Sundrum)
KK graviton at the 14 TeV LHC and a future 100 TeV hadron collider, assuming that the SM fields propagate in the 5D bulk.
To study the discovery potential, we considered the KK graviton decay into two $Z$ bosons, one of which decays leptonically while the other decays invisibly into neutrinos.  Our analysis is then complementary to that in Ref.~\cite{Agashe:2007zd}.  We also discussed how the mass measurement of such a KK graviton can be performed using the energy distribution of the reconstructed $Z(\to \ell^+\ell^-)$, once a discovery is made.

Since the KK gravitons of interest are heavy, each decay product is typically emitted with a large boost. This gives rise to large transverse momenta and highly collimated leptons.  Hence, $\misse$ and $\Delta R_{\ell\ell}$ cuts can be efficient in rejecting the associated SM background. With these cuts, the reach of the 14 TeV LHC for the first KK graviton is 2 TeV (2.6 TeV) with $c\equiv k/\mP=2$ and an integrated luminosity of
300 $\text{fb}^{-1}$ (3 $\text{ab}^{-1}$).  At a future $\sqrt{s}=100$ TeV $pp$ collider,
the relevant reach can extend to $\sim 10$~TeV with 3 $\text{ab}^{-1}$ of data.

The heavy KK gravitons of interest  in our work would be
produced singly and nearly at rest.  On general grounds, one can then expect 
the energy of the visible $Z$ to give a good estimate of the KK graviton mass, 
in the absence of full kinematic information due to invisible neutrinos.      
To find the KK mass, we adapted a novel method, proposed in Ref.~\cite{Agashe:2012bn}, that 
employs the energy distribution of only the visible 
decay product ($Z\to \ell^+\ell^-$) and is suited for our case.   Since the formalism developed in  Ref.~\cite{Agashe:2012bn}
focused on pair-produced resonances (with non-trivial boosts), we adopted a new ansatz
describing the relevant energy distribution for our study.  As a concrete example, we considered a KK graviton with a mass of 2 TeV, 
assuming $c=2$ and 3 $\text{ab}^{-1}$ of 14 TeV LHC data.  With a sample of
$\ord{100}$ KK gravitons, we found that the mass could be deduced at the $\sim 5\%$ level, and the obtained value is fairly precise and accurate. We also briefly discussed possible complementarity of the mass determination approach in our work to potential alternatives.


\section*{Acknowledgments}

We thank Kaustubh Agashe, James Gainer, Ian Lewis, Konstantin Matchev, Frank Paige, Gilad Perez, and Hong Zhang for helpful comments and discussions, Roberto Franceschini for extensive discussions on the fitting functions, Hongsuk Kang, Myeonghun Park and Young Soo Yoon for help with fitting and error estimation. We are especially grateful to  Kaustubh Agashe for useful feedback on a draft version of this work, and Oleg Antipin and Tuomas Hapola for help with the CalHEP Model file. The work of C.-Y.~C. and H.~D. is supported in part by the US DOE Grant DE-AC02-98CH10886. D.~K. was supported in part by NSF Grant No. PHY-0652363, and also acknowledges the support from
the LHC Theory Initiative graduate and postdoctoral fellowship (NSF Grant No. PHY-0969510). D.~K. acknowledges the hospitality of Brookhaven National Laboratory during part of this project.

\end{document}